# Stimulated Radiative Molecular Association in the Early Solar System: Orbital Radii of Satellites of Uranus, Jupiter, Neptune, and Saturn

James C. Lombardi Sr. Professor Emeritus,
Physics Department, Allegheny College, Meadville, PA, USA; james.lombardi@allegheny.edu

## ABSTRACT

The present investigation relates the orbital radii of regular satellites of Uranus, Jupiter, Neptune, and Saturn to photon energies in the spectra of atomic and molecular hydrogen. To explain these observations a model is developed involving stimulated radiative molecular association (SRMA) reactions among the photons and atoms in the protosatellite disks of the planets. In this model thermal energy is extracted from each disk due to a resonance at radii where there is a match between the temperature in the disk and a photon energy. Matter accumulates at these radii, and satellites and rings are ultimately formed. Orbital radii of satellites of Uranus, Jupiter, and Neptune are related to photon energies ($E_{PM}$ values) in the spectrum of molecular hydrogen. Orbital radii of satellites of Saturn are related to photon energies ($E_{PA}$ values) in the spectrum of atomic hydrogen. The first hint that such relationships exist is found in the linearity of the graphs of orbital radii of uranian satellites vs. orbital radii of jovian satellites, as well as in the graphs of orbital radii of uranian satellites vs. orbital radii of neptunian satellites. An expression is determined which gives the temperature in protosatellite disks where the evolution of each satellite begins. This expression is used to find temperature distributions in the disks, which are found to be similar to distributions calculated by other investigators.

## 1. INTRODUCTION

Regular satellites of a planet have orbits with small eccentricities and inclinations relative to the planet's equatorial plane. Sections 2.1 and 2.2 introduce interesting relationships among the orbital radii of regular satellites and rings of the giant gaseous planets in the solar system. Subsequent sections present a possible explanation assuming each young gaseous planet (protoplanet) has a protosatellite disk around it with a radially dependent temperature. In this model regular satellites are formed where disk temperatures have specific resonant values. The spectra of atomic and molecular hydrogen and the mechanism of stimulated radiative association of molecules play key roles in the determination of these specific temperature values. Temperature distributions are determined for the protosatellite disks of Uranus, Jupiter, Neptune and Saturn. These distributions are similar to distributions determined by other investigators. Irregular (captured) satellites are not considered.

## 2. RESULTS and DISCUSSION

### 2.1. Comparing Orbital Radii of Satellites of Uranus, Jupiter and Neptune

The present investigation deals with regular satellites, and from this point on the word "satellite" is understood to mean a regular satellite or ring in the systems of Uranus, Jupiter, Neptune, or Saturn. Table 1 contains the names of known satellites and their orbital radii (lengths of the semi-major axis). The radii are designated $R_{Ui}$, $R_{Ji}$, and $R_{Ni}$ for Uranus, Jupiter, and Neptune respectively. In the case of Uranus, satellites with orbital radii less than 59,000 km are not listed but will be considered in section 2.8. For Jupiter and Neptune all of the known regular satellites are listed. Values for the index $i$ are listed in the first column of the table. All satellites and orbital radii in the same row are associated with the same index. By trial and error, it is determined that associating indices and orbital radii in this way allows us to establish some useful relationships. For instance Fig. 1 is a graph of all the points $(R_{Ui}, R_{Ji})$ for which both $R_{Ui}$ and $R_{Ji}$ are given in the table. In other words the first point in Fig. 1 is the point $(R_{U8}, R_{J8})$, where $R_{U8}$ is the orbital radius of Uranus's satellite Belinda and $R_{J8}$ is the orbital radius of Jupiter's satellite Metis. The particular set of pairings of $R_{Ui}$ and $R_{Ji}$ produces a graph which is fitted very well by a straight line. We would not necessarily expect a linear relationship to exist unless there is a hitherto undescribed mechanism that plays a role in satellite creation. No other set of pairings of uranian and jovian satellite orbital radii is found to produce a graph which is fitted so well. For instance, we can establish another set by pairing the orbital radius of Metis in the Jupiter's system with Perdita in Uranus's system (i.e. a shift of one) and then shift all the other orbital radii in Jupiter's system in the same way. This new set is used to make the graph in Fig. 2 which is not fitted well.

Fig. 3 is a graph of the points $(R_{Ui}, R_{Ni})$ from radii in Uranus's and Neptune's systems, and this graph is also fitted very well by a straight line. No other uranian-neptunian set is well-fitted.

The equations of the best-fit lines to the graphs in Figs. 1 and 3 are

$$R_{Ui} = 0.2050\, R_{Ji} + 49150 \text{ km} \tag{1}$$

and
$$R_{Ui} = 0.9850\, R_{Ni} + 13550 \text{ km.} \tag{2}$$

Ring Galle in Neptune's system lies inside the innermost satellite. Its width is large and optical depth low (NASA 2014). Ring Galle is not included in the present analysis.

There are specific ranges of the index $i$ for which orbital radii are observed in the uranian, jovian, and neptunian systems. Note the four holes in these ranges for which there is no satellite nor orbital radius listed in Table 1 (i.e. $R_{U12}$, $R_{N9}$, $R_{N11}$, and $R_{J12}$). We assume these holes may correspond to satellites that have been gravitationally scattered or not yet observed. It is possible to calculate values for the four orbital radii. From Eq. (2) and $R_{N12}$ we calculate $R_{U12} = 117,200$ km. From Eq. (2) and $R_{U9}$ we calculate $R_{N9} = 63,800$ km. From Eq. (2) and $R_{U11}$ we calculate $R_{N11} = 85,500$ km. From Eqs. (1) and (2) and $R_{N12}$ we calculate $R_{J12} = 332,000$ km.

## 2.2. The Orbital Radii of Satellites of Uranus and Saturn and the Spectra of Molecular and Atomic Hydrogen

The purpose of this section is to show that the orbital radii of satellites of Uranus and Saturn are related to the spectra of molecular and atomic hydrogen. Table 2 includes the orbital radii of satellites of Uranus with indices $i = 1 - 17$ excluding $R_{U12}$, the orbital radius of an unobserved satellite mentioned above. Each of the radii in Table 2 is paired with a particular photon energy ($E_{PM}$) in the spectrum of molecular hydrogen as discussed below. Also included in Table 2 are midplane temperature values ($T_m$'s) in Uranus's protosatellite disk that are also discussed below.

The photon energies of Table 2 are the same as those observed in spectra of nebulae. For example, Martini et al. (1999) have measured the spectra of four reflection nebulae, and Draine and Bertoldi (1996) have studied the spectrum from NGC 2023, one of the nebulae investigated by Martini et al. A paper by Black and van Dishoeck (1987) concerning the fluorescent excitation of interstellar $H_2$ includes a list of wavelengths in the spectrum of $H_2$. From these three papers it is seen that the spectrum of NGC 2023 in the region 1.80-2.40 μm is dominated by peaks corresponding to S-branch transitions in $H_2$. The range of $E_{PM}$ values listed in Table 2 corresponds to this region. None of the $E_{PM}$ values in the region has been omitted except for three that are associated with peaks that are barely apparent in the NGC 2023 spectrum. A discussion concerning the particular pairings of orbital radii with $E_{PM}$ values is found in section 3, the final section of this paper. Suffice it to say at this point we are assuming the radiation detected in the spectrum of NGC 2023 is similar to the radiation that interacted with Uranus's protosatellite disk during the early stages of satellite evolution.

Table 3 is similar to Table 2 in that orbital radii are again associated with a spectrum of photon energies. In this case the orbital radii, $R_S$ values, belong to satellites of Saturn while the photon energies, $E_{PA}$ values, are in the spectrum of atomic hydrogen. The following formula holds for the $E_{PA}$ values, with energy in the units of cm$^{-1}$ (i.e. units with $hc = 1$, where $h$ is Planck's constant and $c$ is the speed of light).

$$E_{PA}(n_f, n_i) = 109737 \text{ cm}^{-1} \ (1/n_f^2 - 1/n_i^2), \ (n_f = 1,2,3\cdots, \ n_i = (n_f+1),(n_f+2)\cdots), \qquad (3)$$

where $n_f$ and $n_i$ are the quantum numbers for the final and initial states associated with the emission of a photon from a hydrogen atom. Photon energies that have the same $n_f$ are all in the same series with the series limit $E_p(n_f, \infty)$.

Table 3 includes a list of photon energy $E_{PA}(n_f, n_i)$ values in a particular range of the atomic hydrogen spectrum. Also listed are quantum numbers ($n_f$ and $n_i$), each corresponding to a specific $E_{PA}(n_f, n_i)$. The listed energies make up a set with each energy in the set paired with an orbital radius, $R_S$ value (NASA 2014), in the satellite system of Saturn. Within the range of $E_{PA}$ values covered by Table 3 there are infinitely many in the series characterized by $n_f = 8$ and 9. We give several of these energies as well as the series limits $E_{PA}(8, \infty)$ and $E_{PA}(9, \infty)$. The unlisted energies leading up to the series limits are responsible for the A and G rings of Saturn. The particular pairings are established by associating the series limit $E_{PA}(8, \infty)$ with the radius of the inner edge of the A ring and the series limit $E_{PA}(9, \infty)$ with the radius of the inner edge of the G ring. After these two pairings are made, all the remaining $E_{PA}$'s in Table 3 are then systematically paired with orbital radii. Further discussion concerning Saturn's rings is in section 2.11. Suffice it to say now that in the present model *the rings of Saturn are related to the closely spaced photon energies near series limits in the hydrogen spectrum*. The photon energies $E_{PA}(6,9)$, $E_{PA}(7,14)$ and $E_{PA}(7,15)$ have values that fall in the range covered by Table 3. However, these energies correspond to orbital radii within Saturn's rings, and it is not possible to associate them with any particular ring features. Therefore, they are not included in the table nor are they included in the analysis.

Next, for the purpose of establishing a useful relationship below, we subtract 3090 cm$^{-1}$ from the $E_{PM}$ values and add 46,700 km to the orbital radii given in Table 3 for Uranus's satellites, i.e.

$$E'_{PM} = E_{PM} - 3090 \text{ cm}^{-1} \tag{4}$$

and

$$R'_{U} = R_{U} + 46700 \text{ km.} \tag{5}$$

Tables 2 and 3 and the shifted coordinates for Uranus's satellites from Eqs. (4) and (5) are used to make the two graphs in Fig. 4 for photon energy distributions associated with Uranus and Saturn. Because satellites are formed in protosatellite disks, the excellent overlap of the two distributions suggests that a connection exists between Uranus's protosatellite disk and photons in the spectra of molecular hydrogen and also between Saturn's protosatellite disk and the spectrum of atomic hydrogen. In later sections we consider why the overlap also indicates that the protosatellite disks of Uranus and Saturn have temperature distributions with similar shapes over an extensive radial distance at the time the evolution of their satellites begins.

A central thesis of the present investigation proposes that the irradiation of protosatellite disks plays an important role in triggering the formation of satellites. If this is true, then the overlap of the graphs in Fig. 4 and the linearity of Eqs. (1) and (2) indicates the protosatellite disks of Uranus, Jupiter, and Neptune are all affected by radiation from molecular hydrogen.

### 2.3. Stimulated Radiative Molecular Association

Radiative molecular association is an important cooling mechanism. Examples of such processes are

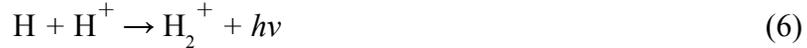
$$H + H^+ \rightarrow H_2^+ + h\nu \tag{6}$$

and

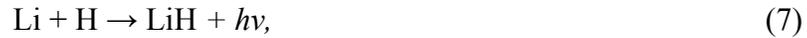
$$Li + H \rightarrow LiH + h\nu, \tag{7}$$

where $h\nu$ designates a photon emitted in the reaction (Dalgarno and Lepp 1987, Gianturco et al. 1996). In the early universe, reactions like these played an important role in the formation of galaxies (Lepp et al. 1998). These reactions cause cooling because in each case a photon is created at the expense of the kinetic energy (due to thermal energy) lost by atoms or ions in the process of forming molecules. Furthermore, cooling continues if, after being formed, the molecules are repeatedly excited into rotational and vibrational levels by collisions with other atoms or molecules and then de-excited by photon emission (Tegmark et al. 1997).

The formation of $H_2$ cannot follow from the collision of two hydrogen atoms because of the lack of a dipole moment in $H_2$. Also because the hydrogen atom has a high ionization energy, the reaction given by Eq. (6) can only be important at temperatures above about 5000 K. This temperature is well above the temperature of a protosatellite disk and so any reaction involving $H^+$ is not important in the present investigation.

If the reaction involves stimulation by a photon, it is called stimulated radiative molecular association (SRMA). An example of this is

$$Li + H + h\nu \rightarrow LiH + 2h\nu \qquad (8)$$

where $h\nu$ on the left side designates the stimulating photon and $2h\nu$ on the right designates both the stimulating and stimulated photons of equal energy (Stancil and Dalgarno 1997). In the present model the reaction given by Eq. (8), or a similar one involving formation of a different molecule, played a key role in the evolution of satellite systems as is discussed below.

In the SRMA calculations of Stancil and Dalgarno (1997) a black-body radiation field impinges upon a gas consisting of Li and H. In the present investigation we assume a discrete spectrum, rather than a continuous black-body spectrum, is causing SRMA in protosatellite disks. As mentioned above, it seems possible that the evolution of the satellite systems of Uranus, Jupiter, Neptune, and Saturn are related to the spectra of molecular and atomic hydrogen. The source of the radiation is not determined here, but it likely originates in the protosun or its protoplanetary disk or in the protosatellite disk itself.

Consider again the SRMA given by Eq. (8). If $E_p$ is the energy of the created photon, then

$$E_p = E_b + \Delta KE, \qquad (9)$$

where $E_b$ is the binding energy of the state in which the LiH molecule is formed and $\Delta KE$ is the kinetic energy (due to thermal energy) that the Li and H system loses during the reaction. We assume $\Delta KE$ is equal to a constant times the temperature at the position in the disk where the SRMA reaction is taking place. The constant is determined in section 2.4.

We note that there are two types of spectra that play roles in this model. One corresponds to the radiation that interacts with the disk (emitted from atomic or molecular hydrogen) and the other is the spectrum of possible levels (each with a different binding energy $E_b$) in which a molecule (possibly LiH) is formed.

In the present model, *a discrete spectrum of photons cools a disk at resonant radii and temperatures where Eq. (9) is satisfied, thus creating relatively cool rings centered at the planet's center. These rings ultimately become regions where matter collects. Therefore satellite orbital radii depend on photon energies.*

### 2.4. Resonance Rings

Although there are many binding energies $E_b$ associated with molecules like LiH, we now propose a condition of resonance in the protoplanetary disk that causes molecules to be formed from only a limited number of binding energies. For the sake of simplifying our discussion we take Eq. (8), involving Li and H, as the reaction that played the important role in the creation of satellite systems even though the precise SRMA is not determined in the present investigation. We make the following hypotheses concerning the collision of a Li atom with a H atom in order to explain the satellite systems that are observed.

1. Eq. (8) involves photons stimulating more photons. Therefore we hypothesize resonance occurs as in a chain reaction.

2. Resonance is created in rings with the following characteristics. Photons that produce many SRMA reactions initially travel tangent to the ring. As they travel they initiate similar reactions and produce photons with the same energy traveling in the same direction establishing the chain reaction. Because of the finite width of the ring, each photon travels a considerable distance before it is not travelling essentially tangent to the ring. These photons all have an energy that satisfies Eq. (9). A *resonance ring* ultimately becomes a region where matter collects and a satellite accretes from the matter.

3. Because resonance causes amplification of the most likely situations, we need to consider rate coefficients for SRMA associated with Eq. (8). Each vibrational energy level of LiH has many rotational components with each component having its own $E_b$ value and rate coefficient. The component (rotational level) with the maximum rate coefficient among all those within the same vibrational level is a candidate for $E_b$ in Eq. (9).

4. Once resonance begins, the temperature and pressure tend to drop within a resonance ring. Perhaps transfer of matter and heat into the ring from adjacent regions helps stabilize the temperature and resonance persists. Or, perhaps the temperature does in fact drop but resonance is strong enough for it to persist. After all, at any temperature there is a distribution of kinetic energies among atoms and molecules in the gas. Therefore there can always be some atoms colliding with kinetic energies that satisfy Eq. (9).

5. $\Delta KE$ in Eq. (9) is equal to a constant C times 2<KE>, where <KE> is the average kinetic energy of an atom in the gas of the disk. C is the fraction of kinetic energy lost by the two atoms during the association process.

Let's consider the value of C. In a SRMA reaction, two atoms collide to form a molecule. Consider the lines of motion of the atoms before the collision occurs. The most probable situation corresponds to the angle being near 90º between these lines. For this case mechanics gives C = 0.5 for any mass ratio of the colliding atoms. Again, as in point 3 above, resonance causes the amplification of the most likely situation. And so we take C to be 0.5 in the determination of $\Delta KE$:

$$\Delta KE = 0.5(2<KE>) = <KE>. \qquad (10)$$

*2.5 Temperature Distributions in Protosatellite Disks*

In the following discussion we assume that SRMA reactions that initiate the evolution of satellites occur near the midplane of a protosatellite disk and we define $T_m$ as the midplane temperature. Because of the symmetry of the disk, we assume $T_m$ is only radially dependent. We also use the following relationship derived from statistical mechanics to relate $T_m$ to <KE>.

$$<KE> = (3/2)k_B T_m, \qquad (11)$$

where $k_B$ is the Boltzmann constant.

From Eqs. (9), (10) and (11) we derive

$$(3/2\ k_B T_m) = E_{PS} - E_b$$

We want to express $E_{PS}$ and $E_b$ in the units of wavenumbers (cm$^{-1}$) so the above relationship becomes

$$(3/2\ k_B T_m) = hc(E_{PS} - E_b),$$

where h is the Planck constant and c is the speed of light. So

$$T_m = 2/3\ hc/k_B\ (E_{PS} - E_b), \qquad (12)$$

where hc/k$_B$ = 1.439 cm kelvin.

We have that there are linear relationships among the orbital radii of satellites of Uranus, Jupiter, and Neptune (Eq. (1) and (2)). There is also excellent overlap between photon energy distributions for Uranus and Saturn (Fig. 4). These facts indicate that satellite migration is either not very large, or is uniform among the satellites. Therefore, if we use Eq. (12) to determine $T_m$ values in a protosatellite disk where primordial rings are established, and graph $T_m$ vs. orbital radii, the graph will be closely related to the temperature distribution in the disk at the time when the evolution of satellites begins.

Henceforth, TD stands for the *midplane temperature distribution or portion of a temperature distribution in the protosatellite disk of a planet where and when the evolution of its satellites begins.*

### 2.6. $E_b$ and Stimulated Radiative Association

In order to consider TD's of Uranus, Jupiter, and Neptune, we need to determine the value of $E_b$ in Eq. (12). For a particular SRMA, the rotational state with the maximum rate coefficient among all those within the same vibrational level is a candidate for $E_b$. It would be very difficult if not impossible to determine $E_b$ by considering the many possible SRMA reactions and their many rotational-vibrational levels. Therefore we take $E_b$ to be an adjustable parameter and vary it so that Eq. (12) fits a previously determined temperature distribution as discussed below.

Fig. 5 includes a reproduction of a temperature distribution (thin line) calculated by Mousis (2003) for a thin gaseous "water-rich subnebula" circulating around Uranus. The scale for these radii is the equatorial radius of Uranus, $R_{Uranus}$ = 25,559 km. Mousis (2003) actually has a series of distributions, each for a different time after Uranus experienced a single giant impact. The Mousis (2003) distribution reproduced in Fig. 5 is the earliest distribution, corresponding to $10^4$ years after the impact. We determine various trial uranian TD's, each with a different value of $E_b$ in Eq. (12), in an attempt to match the Mousis distribution, by using the orbital radii ($R_{Ui}$) and photon energies ($E_{PM}$ values) found in Table 2. Midplane temperatures ($T_m$'s) are calculated by using $E_{PM}$'s for $E_{PS}$ in Eq. (12), and trial TD's are constructed by plotting $T_m$ vs. $R_{Ui}/R_{Uranus}$. Two distributions found in this way are represented by the filled and open circles in Fig. 5, calculated with $E_b$ equal to 2500 and 2700 cm$^{-1}$ respectively. After considering these results, we take $E_b$ to be 2500 cm$^{-1}$ in subsequent calculations. The $T_m$ values found using $E_b$ = 2500 cm$^{-1}$ appear in Table 2 and the distribution in Fig. 5 made with these $T_m$ values is taken to be the TD of Uranus.

In section 2.9 below the $E_b$ for the TD of Saturn is determined to be 945 cm$^{-1}$ in a similar data fitting process.  This value and the 2500 cm$^{-1}$ value found for Uranus are consistent with binding energies of states of LiH that have high radiative molecular association rate coefficients.  The rate coefficients are highest for the molecule being formed in states with vibrational quantum number v equal to 18 (Gianturco and Giorgi 1997).   Some of the binding energies for purely vibrational states with v near 18 are $E_b$ (v =16) = 2936 cm$^{-1}$, $E_b$ (v =18) = 1679 cm$^{-1}$ and $E_b$ (v =20) = 705 cm$^{-1}$ (Shi et al 2013).  Besides the purely vibrational states, there are many levels with  rotational components.  All of these have different rate coefficients and different binding energies.  The $E_b$ values determined in the present analysis for Uranus and Saturn (945 cm$^{-1}$ and 2500 cm$^{-1}$ respectively) are therefore consistent with binding energies of accessible states in LiH.  However this does not prove conclusively that LiH is the molecule that played the key role in the SRMA processes that initiated satellite formation in the protosatellite disks of Saturn and Uranus.

Fig. 5 indicates Uranus's TD agrees with the Mousis distribution at t = 10$^4$ years for radii between about 2.3 $R_{Uranus}$ and 20 $R_{Uranus}$.  Later we see these distributions do not agree for smaller radii.  But we will also see the uranian TD found from Eq. (12) does have a shape which is similar to a temperature distribution previously calculated for the solar nebula even for small radii.  It is not possible to use Eq. (12) to suitably fit any of the Mousis (2003) distributions other than the one corresponding to 10$^4$ years after the impact.

## 2.7. TD's of Uranus, Jupiter and Neptune

We now consider the relationship among TD's of Uranus, Jupiter, and Neptune.  Eq. (1) is a transformation equation for the radial coordinate of Jupiter's protosatellite disk and satellite system.  The first constant on the right side of the equation scales the radial coordinate, and the second constant shifts the origin such that orbital radii of Jupiter's satellites have values that are close to the orbital radii of Uranus's satellites.  Eq. (2) transforms Neptune's radial coordinate in the same way.  These linear relationships exist because the TD's of the three planets have similar shapes.  The present investigation does not explain why these distribution are similar but simply accepts that they are.

It is possible to construct a composite TD for Uranus, Jupiter, and Neptune.  For this the orbital radii of Jupiter's and Neptune's satellites in Table 1 are transformed using Eqs. (1) and (2) and midplane temperatures ($T_m$ values) are taken from Table 2.  The values of $i$ in Tables 1 and 2 help match up $T_m$'s with the appropriate satellite orbital radii for Jupiter and Neptune. Table 4 lists the $T_m$'s for the three planets, the orbital radii of Uranus's satellites, and the transformed orbital radii $R_{Jti}$ and $R_{Nti}$ for Jupiter's and Neptune's satellites respectively.  The orbital radii are given both in km and in units of $R_{Uranus}$.  This table is used to construct Fig. 6a, the composite TD for the satellites of Uranus, Jupiter and Neptune.  Notice how well points for Jupiter and Neptune overlap Uranus points.  The overlap is so good that it is helpful to use the scheme on the left side of the figure to tell if each position on the graph corresponds to one, two, or three points.  This point by point agreement supports one of the main theses of this paper: *the same mechanism initiated the evolution of the satellites in the giant gas planets*.  One of Neptune's points plotted near the coordinate 5 $R_{Uranus}$ stands out as not overlapping with a Uranus or a Jupiter point.  This position on the graph corresponds to satellites of Uranus and Jupiter that have not been observed, but for which orbital radii are calculated in section 2.1.

Alibert et al. (2005) have calculated temperature distributions for various times during the evolution of Jupiter's disk. In order to easily compare the present results to Fig. 6 in Alibert et al. (2005), the data in the composite TD in Fig. 6a are replotted in the log-log graph of Fig 6b. The composite TD is very similar in shape and has nearly the same temperature values as the distributions corresponding to early times in Jupiter's disk (Alibert et al. 2005).

## 2.8. The uranian TD and the temperature distribution of the solar nebula

Table 4 has $T_m$ values for Uranus's satellites with orbital radii greater than 59,000 km, i.e. Bianca and beyond. Table 5 extends these data inward from Bianca to smaller radii so that the remainder of Uranus's satellites are included. Figs. 7-9 will help explain why some orbital radii in Table 5 correspond to the same $E_{PM}$ and the same index $i$ as in Table 4. First consider Fig. 7. It is a logarithmic line graph of the orbital radii of Uranus's satellites from Ring 6 to Portia. The square blocks designating the points have been made large enough so that most of them overlap. On the right side of the graph the larger space and two smaller spaces to the right and left of it indicate that this region of orbital radii corresponds to a broad peak in the TD with its maximum just to the left of Bianca. Because Bianca has $i = 1$, the satellite just inward from Bianca also has $i = 1$, and is paired with the same $E_{PM}$ and $T_m$ values as Bianca. The other space on the left side of Fig. 7 corresponds to a valley in the TD.

Table 5 includes components of ring η labeled $η_1$ and $η_2$. The orbital radius of component $η_2$ is derived from data in French et al. (1991) especially pages 371, 372 and 380. The authors write of ring η's two distinct components. One is a "sharp inner feature" assigned the orbital radius of 47,176 km and the other is a "component which extends ~ 55 km exterior to the sharp feature." The model put forth in the present paper predicts there should be two close rings or satellites in the vicinity of ring η. We assume these to be associated with the two observed components of η and derive the approximate orbital radius of $η_2$ to be the sum of 47,176 and 55 km.

Tables 4 and 5 are used to construct the complete TD for the disk of Uranus extending over the region of the planet's regular satellites. This TD is shown in Fig. 8. Given that processes which govern the behavior of protosatellite disks and protoplanetary disks are similar (Mousis et al. 2002, Mousis 2003, and Mousis and Alibert 2006), we now compare Fig. 8 to temperature distributions calculated by Lin and Papaloizou (1985, hereafter L&P) for the solar nebula. L&P perform a self-consistent global analysis of the solar nebula to characterize its properties. Fig. 18 of L&P contains two sets of graphs showing midplane temperature distributions of the solar nebula for different times during what are called upward and downward transition waves. We find the TD in Fig. 8 is similar in shape to the L&P solar nebula distributions. Fig. 9 is a reproduction of two L&P distributions, one for shortly after onset of the downward transition and one near the end of of the transition. Both of these distribution are peaked and have characteristics similar to those of the TD in Fig. 8. To the right of the peak in the L&P distribution at the onset of the transition, there is a shoulder and then a long negatively sloped tail just as in the TD of Fig. 8. Also, in the L&P distribution near the end of the downward transition, the temperature distribution goes down to the left of the peak and then up again just as in the TD in Fig. 8.

### 2.9. The saturnian TD

Mousis, Gautier, and Bockelee-Moran (2002) (referred to as MGB) use an analytical evolutionary model to determine temperature distributions of Saturn's subnebula (protosatellite disk) throughout a large portion of its lifetime. Their sixth and last distribution corresponds to $5 \times 10^6$ years. The orbital radii of the saturnian satellites and the photon energies $E_{pA}$ found in Table 3 are previously used in the present investigation to illustrate the overlap of the saturnian and uranian photon energy distributions in Fig. 4. The same data are now used along with Eq. (12) to fit the MGB, $t = 5 \times 10^6$ year temperature distribution by varying $E_b$. This is the same procedure used to fit the Mousis (2003) temperature distribution with the uranian TD that is detailed in section 2.6. A good fit is found when $E_b$ has the value of 945 cm$^{-1}$. Because we are unable to suitably fit any of the MGB distributions corresponding to other times, we conclude that the fit to the $t = 5 \times 10^6$ year distribution is the saturnian TD. The MGB distribution is given by the solid curve in Fig. 10. The individual data points represent the saturnian TD. Radii in Fig. 10 are measured in units of Saturn's equatorial radius $R_{Sat} = 60,268$ km.

Hyperion is the last satellite listed in Table 3. It is the most distant satellite from Saturn with an orbital inclination and orbital eccentricity that classify it as regular. However Hyperion has a highly irregular shape which is atypical for a regular satellite. It is believed that Hyperion is what remains from a much larger satellite that experienced a catastrophic impact (Farinella et al. 1997). Most of the many fragments created by the collision were swept up or away by Titan. However after impact some fragments were in a 4:3 mean-motion resonance with Titan, preventing close encounters of the fragments with Titan. The present day Hyperion is believed to have accreted from these fragments (Farinella et al. 1997). Therefore Hyperion's orbital radius is different from the orbital radius of the satellite that originally orbited beyond Titan. It is possible to estimate the orbital radius of the original satellite by using the $T_m$ value associated with Hyperion in Table 3. This value is $T_m = (7.3 \pm 3)$ K with the uncertainty determined by considering the range of reasonable fits of Eq. (12) to the MGB curve in Fig. 10. The temperature of $(7.3 \pm 3)$ K on the MGB curve corresponds to an orbital radius on the order of 100 $R_{Sat}$, a few times larger than the orbital radius of the present day Hyperion.

### 2.10. The composite TD for Uranus, Jupiter, Neptune and Saturn

For the purpose of illustrating the similarity of the TD's for the disks of Uranus, Jupiter, Neptune, and Saturn we now construct a composite TD in which the four TD's overlap. First consider Eq. (5) once more. Previously this relationship was used to adjust the radial scale of the photon energy distribution associated with Uranus so it overlapped with the photon distribution associated with Saturn. Now we are transforming in the other direction. That is, we want to rearrange Eq. (5) so it has a form analogous to Eqs. (1) and (2). The new relationship is

$$R_U = R_S - 46700 \text{ km.} \qquad (13)$$

We use Eq. (13) to transform the orbital radii of Saturn's satellites that are listed in Table 3. These transformed coordinates along with the coordinates and transformed coordinates found in Tables 4 and 5 are used to plot the four TD's.

But we also need to make a shift along the temperature axis. Overlap of the four TD's is achieved by subtracting 1480 K from the temperatures in the TD's of Uranus, Jupiter and Neptune in Table 4. The composite TD is given in Fig. 11. The open circles represent the TD for Saturn and the filled circles represent the TD's for Uranus, Jupiter and Neptune. There is close overlap of the filled and open circles. Also the overlap of the Uranus, Jupiter and Neptune TD's is so exact that all points with the same $i$ fall on one another. The numbers 2 or 3 next to some of the filled circles indicate where 2 or 3 points belonging to different TD's overlap. Filled circles without a number correspond mainly to the uranian TD. Again, the present investigation indicates that temperature distributions in protosatellite disks have similar shapes where and when the evolution of satellites begins but does not explain why.

## 2.11. Saturn's Rings

In section 2.2 and Table 3 the series limit $E_{PA}(8,\infty)$ is paired with the inner edge of Saturn's A ring. We now consider this relationship. In this discussion $E_{PS} = E_{PA}$, i.e. the stimulating photon has an energy equal to the energy in the spectrum of atomic hydrogen. Wherever $E_{PA}$ satisfies Eq. (12) in a protoplanetary disk, resonance is established and thermal energy is extracted. Because resonance rings have finite widths, near the series limit $E_{PA}(8,\infty)$ many closely spaced photon energies create overlapping resonance rings and a circular band of resonance is created. Cooling occurs there, matter collects and ultimately the A ring is formed. Within the region of the A ring, as the radial coordinate decreases, the temperature of the disk and therefore the values of $E_{PA}$ increases. This means that $E_{PA}(8,\infty)$ correspond to the inner edge of the ring. Within the region of the A ring, as the radial coordinate increases, the radii of component resonance rings separate more and more and eventually the resonance rings do not overlap. The radial coordinate where overlap ceases defines the outer edge of the A ring. Notice in Table 3 the outer edge radius of the A ring is nearly equal to the orbital radius of Daphnis, the satellite associated with photon energy $E_{PA}(8, 23)$. There is no satellite paired with $E_{PA}(8, n_i \geq 24)$. We therefore conclude that the outermost resonance ring that contributes to the A ring is associated with $E_{PA}(8, 24)$.

Also in section 2.2 and Table 3, $E_{PA}(9,\infty)$ is paired with the inner edge of the G ring. Most of the discussion above that applies to the A ring also applies to the G ring. However beyond the G ring there are no satellites that are associated with $E_{PA}$ values with $n_f = 9$. A possible explanation for this involves the emission intensities in atomic hydrogen. These intensities are generally higher for lower values of $n_f$ (Wiese and Fuhr 2009). Possibly the intensities for which $n_f < 9$ are strong enough to create isolated resonances rings. But for $n_f \geq 9$, intensities are not strong enough to cause resonance unless resonance rings spatially overlap and mutually support each other as within the region of the G ring.

The present investigation does not include analyses of Saturn's B, C, D and E rings. The B, C, and D rings do not exist in the region of Saturn's TD derived in this investigation. Also, it is believed that the E ring is made of small dust particles that originate from geophysical activity on the satellite Enceladus (Spahn et al. 2006).

## 3. FURTHER DISCUSSION

The present investigation is initiated from two key observations. The first is that simple relationships exist among the orbital radii of satellites of Uranus, Jupiter and Neptune, as seen in Figs. 1 and 3. The second is the overlap of the photon energy distributions associated with the orbital radii of the uranian and saturnian satellites, as seen in Fig. 4. To explain these observations a resonance model is assumed that involves stimulated radiative molecular association (SRMA) of molecules. This reaction is appealing for two main reasons. First, it explains why the spectra of atomic and molecular hydrogen are essential to the construction of Fig. 4. Second, because SRMA involves photons stimulating more photons it provides a means for resonance that creates features of satellite systems such as nearly circular orbits and well-defined inner edges of Saturn's A and G rings. If only the linear relationships in Figs. 1 and 3 are observed and not the overlap in Fig. 4, this would be interesting but there would be no way to justify using SRMA to explain the observation. If the overlap in Fig. 4 is observed but not the linear relationships, then there would be no way to tie together all the TD's as in Fig. 11.

Interestingly the coefficients of $R_U$ and $R_S$ in Eq. (13) are unity. Why should the TD's of the disks of Uranus and Saturn have the same radial scale? This also is nearly true for disks of Uranus and Neptune: consider Eq. (2). On the other hand Eq. (1) shows the radial scale for the TD of Uranus is about one fifth the radial scale for the disk of Jupiter: consider Eq. (1).

To establish the TD in Saturn's protosatellite disk the series limits $E_{PA}(8,\infty)$ and $E_{PA}(9,\infty)$ are paired with the inner edges of the A and G rings. If any other scheme is used, then the saturnian TD doesn't overlap the uranian TD so well. Furthermore, the connection between the close photon energies near series limits and the rings of Saturn helps explain the existence of the rings.

In section 2.2 it was mentioned that this last section would contain a discussion concerning the pairings of orbital radii with $E_{PM}$ values in Tables 2 and 5. Table 5 contains the orbital radii of the rings of Uranus. Most of these rings were observed in 1977 and reported by Elliot (1979). Figure 8 of Elliot's paper shows the occultation of a star's light by the uranian ring system. The author of the present paper noticed a correspondence between the positions of the dips in the Elliot figure and the peaks in the $H_2$ spectrum in Fig. 1 of Martini et al. (1999). This agreement leads to the pairings in Table 5 and then to the pairings in Table 2. But ultimately, justifications for all the orbital radii/photon energy pairings in this investigation are

1. the ability to suitably fit Uranus's TD to the previously determined temperature distribution in Uranus's subnebula (Fig. 5),
2. the ability to suitably fit Saturn's TD to the previously determined temperature distribution in Saturn's subnebula (Fig. 10)
3. the similarity between the composite TD of Uranus, Jupiter and Neptune with the previously determined temperature distribution in Jupiter's subnebula (Fig 6b and Alibert (2005),
4. the similarity of Uranus's TD with temperature distributions in the solar nebula (Figs 8 and 9), and
5. the overlap of the TD's of Uranus, Jupiter, Neptune and Saturn (Figs. 6a and 11).

## Acknowledgments

The author thanks James Lombardi Jr. and Jerry Staub for helpful suggestions on the manuscript for this paper, and Allegheny College for use of computer facilities.

**Table 1.** Orbital radii of satellites and rings of Uranus, Neptune and Jupiter.

| $i$ | Uranian Satellites | $R_{Ui}$(km)[a] | Neptunian Satellites | $R_{Ni}$(km)[a] | Jovian Satellites | $R_{Ji}$(km)[a] |
|---|---|---|---|---|---|---|
| 1 | Bianca | 59170 | | | | |
| 2 | Cressida | 61780 | Naiad | 48227 | | |
| 3 | Desdemona | 62680 | Thalassa | 50075 | | |
| 4 | Juliet | 64350 | Despina | 52526 | | |
| 5 | Portia | 66090 | Ring LeVerrier | 53200 | | |
| 6 | Rosalind | 69940 | Ring Arago | 57200 | | |
| 7 | Cupid | 74800 | Galatea | 61953 | | |
| 8 | Belinda | 75260 | Ring Adams | 62933 | Metis | 128000 |
| 9 | Perdita | 76400 | Not Observed | | Adrastea | 129000 |
| 10 | Puck | 86010 | Larissa | 73548 | Amalthea | 181400 |
| 11 | Mab | 97700 | Not Observed | | Thebe | 221900 |
| 12 | Not Observed | | S/2004 N1 | 105300 | Not Observed | |
| 13 | Miranda | 129390 | Proteus | 117650 | Io | 421600 |
| 14 | Ariel | 191020 | | | Europa | 670900 |
| 15 | Umbriel | 266300 | | | Ganymede | 1070400 |
| 16 | Titania | 435910 | | | Callisto | 1882700 |
| 17 | Oberon | 583520 | | | | |

[a]NASA (2014)

**Table 2.** Orbital radii of Uranian satellites and associated photon energies and midplane temperatures. The column titled $H_2$ Transition identifies the H2 transition that produces the particular photon energy.

| $i$ | Uranian Satellites | $R_{Ui}$(km)[a] | $H_2$ Transition | $E_{PM}$(cm⁻¹)[b] | $T_m$(K)[c] |
|---|---|---|---|---|---|
| 1 | Bianca | 59170 | (2,1) S(7) | 5397.2 | 2779.4 |
| 2 | Cressida | 61780 | (1,0) S(4) | 5285.6 | 2672.3 |
| 3 | Desdemona | 62680 | (9,7) S(1) | 5146.8[d] | 2539.2 |
| 3 | Desdemona | 62680 | (2,1) S(5) | 5141.8[d] | 2534.4 |
| 4 | Juliet | 64350 | (1,0) S(3) | 5108.4 | 2502.3 |
| 5 | Portia | 66090 | (2,1) S(4) | 4989.8 | 2388.5 |
| 6 | Rosalind | 69940 | (1,0) S(2) | 4917.0 | 2318.7 |
| 7 | Cupid | 74800 | (3,2) S(5) | 4841.3 | 2246.1 |
| 8 | Belinda | 75260 | (2,1) S(3) | 4822.8 | 2228.3 |
| 9 | Perdita | 76400 | (1,0) S(1) | 4712.9 | 2122.9 |
| 10 | Puck | 86010 | (2,1) S(2) | 4642.1 | 2055.0 |
| 11 | Mab | 97700 | (3,2) S(3) | 4542.6 | 1959.5 |
| 12 | Not Observed | | (1,0) S(0) | 4497.8 | 1916.6 |
| 13 | Miranda | 129390 | (2,1) S(1) | 4449.0 | 1869.7 |
| 14 | Ariel | 191020 | (3,2) S(2) | 4372.4 | 1796.3 |
| 15 | Umbriel | 266300 | (4,3) S(3) | 4265.2 | 1693.4 |
| 16 | Titania | 435910 | (2,1) S(0) | 4245.2 | 1674.2 |
| 17 | Oberon | 583520 | (3,2) S(1) | 4190.2 | 1621.5 |

[a]NASA (2014)

[b]Black and van Dishoeck (1987)

[c]Midplane temperature $T_m$ values determined using the binding energy $E_b = 2500$ cm⁻¹

[d]The close $E_{PM}$ values of 5141.8 and 5146.8 cm-1 are averaged in the analysis to give a value of $T_m = 2536.8$ K for Desdemona.

**Table 3.** Orbital radii of Saturnian satellites associated with photon energies from Eq. (3). Also included are midplane temperatures which are discussed in section 2.9.

| Saturnian Satellites | $R_s$(km)[a] | $n_f, n_i$ | $E_{P_A}(n_f, n_i)$(cm$^{-1}$) | $T_m$(K) |
|---|---|---|---|---|
| S/2009 S1 | 117000 | 7,16 | 1810.9 | 830.7 |
| IE [b] A ring | 122170 | 8,∞ | 1714.6 | 738.3 |
| Pan | 133583 | 7,13 | 1590.2 | 619.0 |
| Daphnis | 136500 | 8,23 | 1507.2 | 539.3 |
| OE [b] A ring[c] | 136775 | | | |
| Atlas | 137670 | 8,22 | 1487.9 | 520.8 |
| Prometheus | 139350 | 7,12 | 1477.5 | 510.8 |
| F ring | 140180 | 8,21 | 1465.8 | 499.6 |
| Pandora | 141700 | 8,20 | 1440.3 | 475.2 |
| Ep.& Jan.[d] | 151450 | 8,19 | 1410.7 | 446.8 |
| Aegaeon | 167500 | 8,18 | 1376.0 | 413.5 |
| IE G ring | 170000 | 9,∞ | 1354.8 | 393.1 |
| OE G ring | 175000 | | | |
| Mimas | 185200 | 5,6 | 1341.2 | 380.1 |
| Methone | 194000 | 8,17 | 1334.9 | 374.0 |
| Anthe | 197700 | 6,8 | 1333.6 | 372.8 |
| Pallene | 211000 | 7,11 | 1332.6 | 371.8 |
| Enceladus | 238020 | 8,16 | 1286.0 | 327.1 |
| Tethys | 294660 | 8,15 | 1226.9 | 270.4 |
| Dione | 377400 | 8,14 | 1154.8 | 201.3 |
| Rhea | 527040 | 7,10 | 1142.2 | 189.2 |
| Titan | 1221830 | 8,13 | 1065.3 | 115.4 |
| | | 10,∞ | 1097.4[e] | |
| Hyperion | 1481100 | 8,12 | 952.6 | 7.3 |

[a]NASA (2014)

[b]IE & OE stand for inner and outer edge.

[c]The outer edge of the A ring is discussed in section 2.11.

[d]Epimetheus & Janus

[e]This photon series limit is not considered in the analysis because no inner edge of a ring has been observed beyond the orbital radius of Titan.

**Table 4.** Orbital radii and $T_m$ values used to construct overlapping TD's for Uranus, Neptune and Jupiter

| $i$ | $H_2$ Transition | $T_m$(K)[a] | Uranian Sat's | $R_{Ui}$(km)[b] | $R_{Ui}/R_{Uranus}$ | Neptunian Sat's | $R_{Nti}$(km)[c] | $R_{Nti}/R_{Uranus}$ | Jovian Sat's | $R_{Jti}$(km)[d] | $R_{Jti}/R_{Uranus}$ |
|---|---|---|---|---|---|---|---|---|---|---|---|
| 1 | (2,1) S(7) | 2779.4 | Bianca | 59170 | 2.315 | | | | | | |
| 2 | (1,0) S(4) | 2672.3 | Cressida | 61780 | 2.417 | Naiad | 61008 | 2.387 | | | |
| 3 | see note[e] | 2536.8 | Desdemona | 62680 | 2.452 | Thalassa | 62827 | 2.458 | | | |
| 4 | (1,0) S(3) | 2502.3 | Juliet | 64350 | 2.518 | Despina | 65238 | 2.552 | | | |
| 5 | (2,1) S(4) | 2388.5 | Portia | 66090 | 2.586 | Ring LeVerrier | 65902 | 2.578 | | | |
| 6 | (1,0) S(2) | 2318.7 | Rosalind | 69940 | 2.736 | Ring Arago | 69839 | 2.732 | | | |
| 7 | (3,2) S(5) | 2246.1 | Cupid | 74800 | 2.927 | Galatea | 74516 | 2.915 | | | |
| 8 | (2,1) S(3) | 2228.3 | Belinda | 75260 | 2.945 | Ring Adams | 75480 | 2.953 | Metis | 75392 | 2.950 |
| 9 | (1,0) S(1) | 2122.9 | Perdita | 76400 | 2.989 | Not Observed | | | Adrastea | 75597 | 2.958 |
| 10 | (2,1) S(2) | 2055.0 | Puck | 86010 | 3.365 | Larissa | 85927 | 3.362 | Amalthea | 86339 | 3.378 |
| 11 | (3,2) S(3) | 1959.5 | Mab | 97700 | 3.823 | Not Observed | | | Thebe | 94641 | 3.703 |
| 12 | (1,0) S(0) | 1916.6 | Not Observed | | | S/2004 N1 | 117174 | 4.584 | Not Observed | | |
| 13 | (2,1) S(1) | 1869.7 | Miranda | 129390 | 5.062 | Proteus | 129327 | 5.060 | Io | 135580 | 5.305 |
| 14 | (3,2) S(2) | 1796.3 | Ariel | 191020 | 7.474 | | | | Europa | 186687 | 7.304 |
| 15 | (4,3) S(3) | 1693.4 | Umbriel | 266300 | 10.419 | | | | Ganymede | 268584 | 10.508 |
| 16 | (2,1) S(0) | 1674.2 | Titania | 435910 | 17.055 | | | | Callisto | 435106 | 17.024 |
| 17 | (3,2) S(1) | 1621.5 | Oberon | 583520 | 22.830 | | | | | | |

[a] $T_m$ values determined using $E_b$ = 2500 cm$^{-1}$

[b] NASA (2014)

[c] These orbital radii are from NASA (2014) and then transformed using Eq. (2).

[d] These orbital radii are from NASA (2014) and then transformed using Eq. (1).

[e] The $E_{iui}$(cm$^{-1}$) for $i$ = 3 is the average of 5146.8 & 5141.8 cm$^{-1}$ from two $H_2$ transitions, (9,7)S(1) & (2,1) S(5).

This results in a single value of 2536.8 K for $T_m$.

**Table 5.** Orbital radii less than 54,000 km for Uranian satellites and associated photon energies and midplane temperatures.

| $i$ | Uranian Sat's | $H_2$ Transition | $E_{PM}(cm^{-1})$[a] | $T_m(K)$[b] | $R_{Ui}(km)$[c] | $R_{Ui}/R_{Uranus}$ |
|---|---|---|---|---|---|---|
| 8 | Ring 6 | (2,1) S(3) | 4822.8 | 2228.3 | 41837 | 1.637 |
| 9 | Ring 5 | (1,0) S(1) | 4712.9 | 2122.9 | 42234 | 1.652 |
| 10 | Ring 4 | (2,1) S(2) | 4642.1 | 2055.0 | 42571 | 1.666 |
| 10 | Ring α | (2,1) S(2) | 4642.1 | 2055.0 | 44718 | 1.750 |
| 9 | Ring β | (1,0) S(1) | 4712.9 | 2122.9 | 45661 | 1.786 |
| 8 | Ring $\eta_1$ | (2,1) S(3) | 4822.8 | 2228.3 | 47176 | 1.846 |
| 7 | Ring $\eta_2$[d] | (3,2) S(5) | 4841.3 | 2246.1 | 47231 | 1.848 |
| 6 | Ring γ | (1,0) S(2) | 4917.0 | 2318.7 | 47627 | 1.863 |
| 5 | Ring δ | (2,1) S(4) | 4989.8 | 2388.5 | 48300 | 1.890 |
| 4 | Cordelia | (1,0) S(3) | 5108.4 | 2502.3 | 49770 | 1.947 |
| 3 | Ring λ | see note[e] | 5144.3 | 2536.8 | 50024 | 1.957 |
| 2 | Ring ε | (1,0) S(4) | 5285.6 | 2672.3 | 51149 | 2.001 |
| 1 | Ophelia | (2,1) S(7) | 5397.2 | 2779.4 | 53790 | 2.105 |

[a]Black and van Dishoeck (1987)

[b]$T_m$ values determined using $E_b = 2500$ cm$^{-1}$

[c]NASA (2014) except for Ring $\eta_2$

[d]French et al. (1991). Especially pp. 371, 372, and 380.
   Also see the text regarding this component of η.

[e]The $E_{PM}(cm^{-1})$ for $i = 3$ is the average of 5146.8 & 5141.8 cm$^{-1}$

   from two $H_2$ transitions, (9,7)S(1) & (2,1) S(5).

   This results in a single value of 2536.8 K for $T_m$.

Figure 1. The orbital radii of uranian satellites vs. the orbital radii of jovian satellites when the satellites are correctly paired. The linear fit to the data is good.

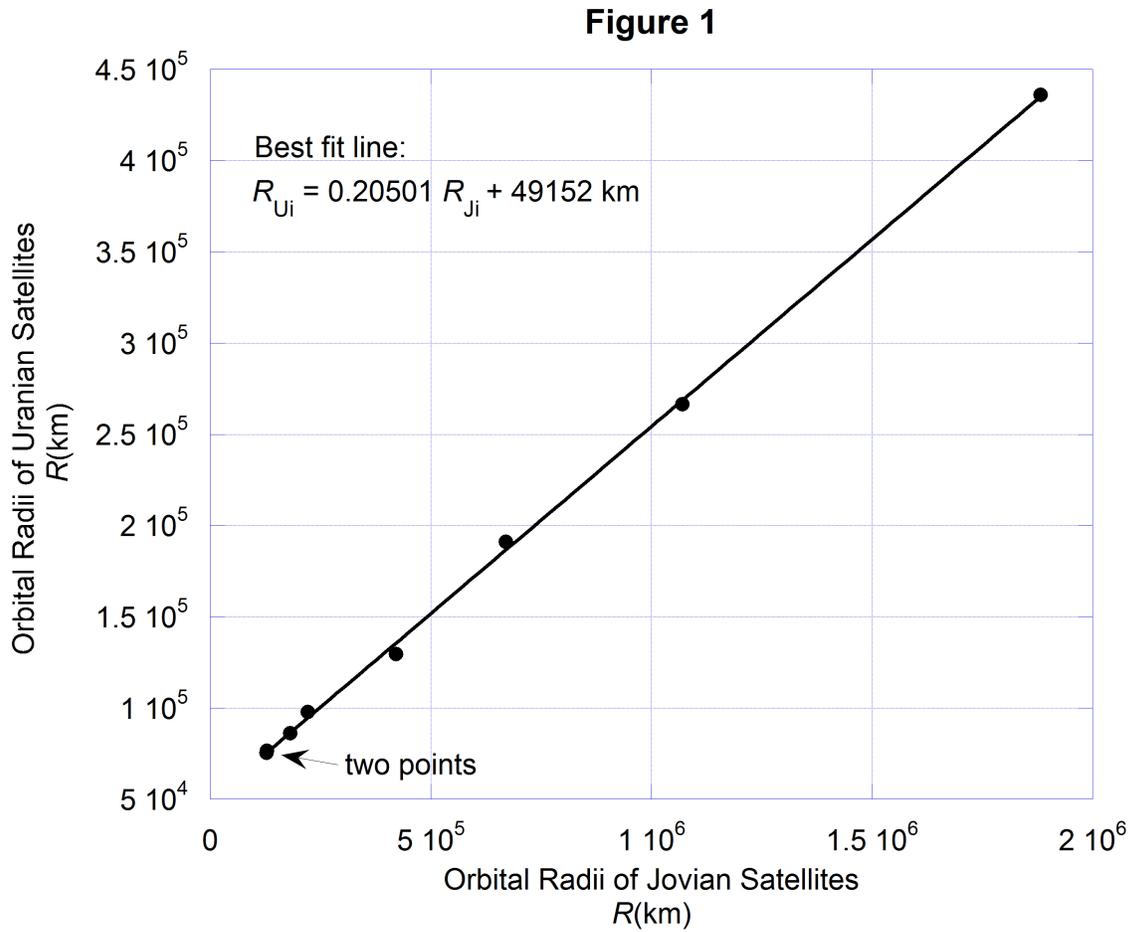

Figure 2.  The orbital radii of uranian satellites vs. the orbital radii of jovian satellites when the
satellites are not paired correctly.  The linear fit to the data is poor.

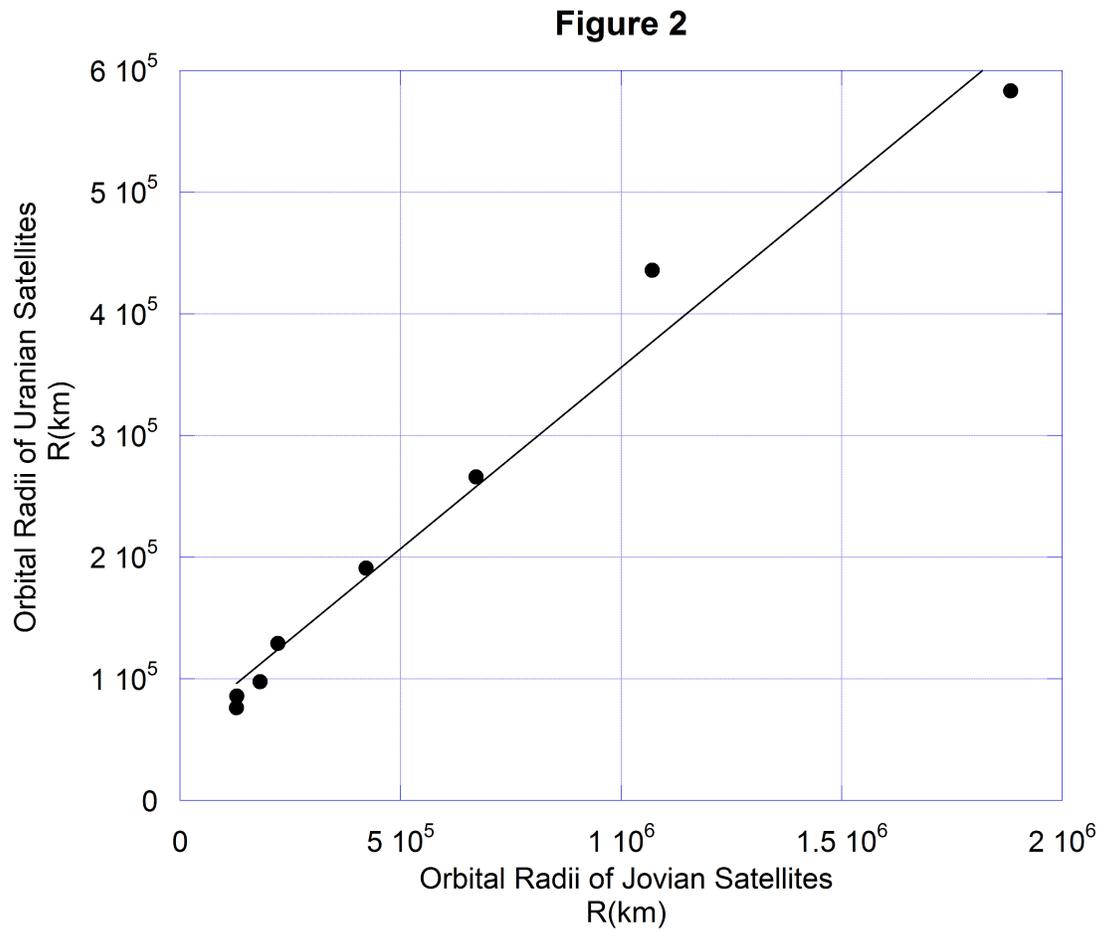

Figure 3.  The orbital radii of uranian satellites vs. the orbital radii of neptunian satellites when the satellites are correctly paired.  The linear fit to the data is good.

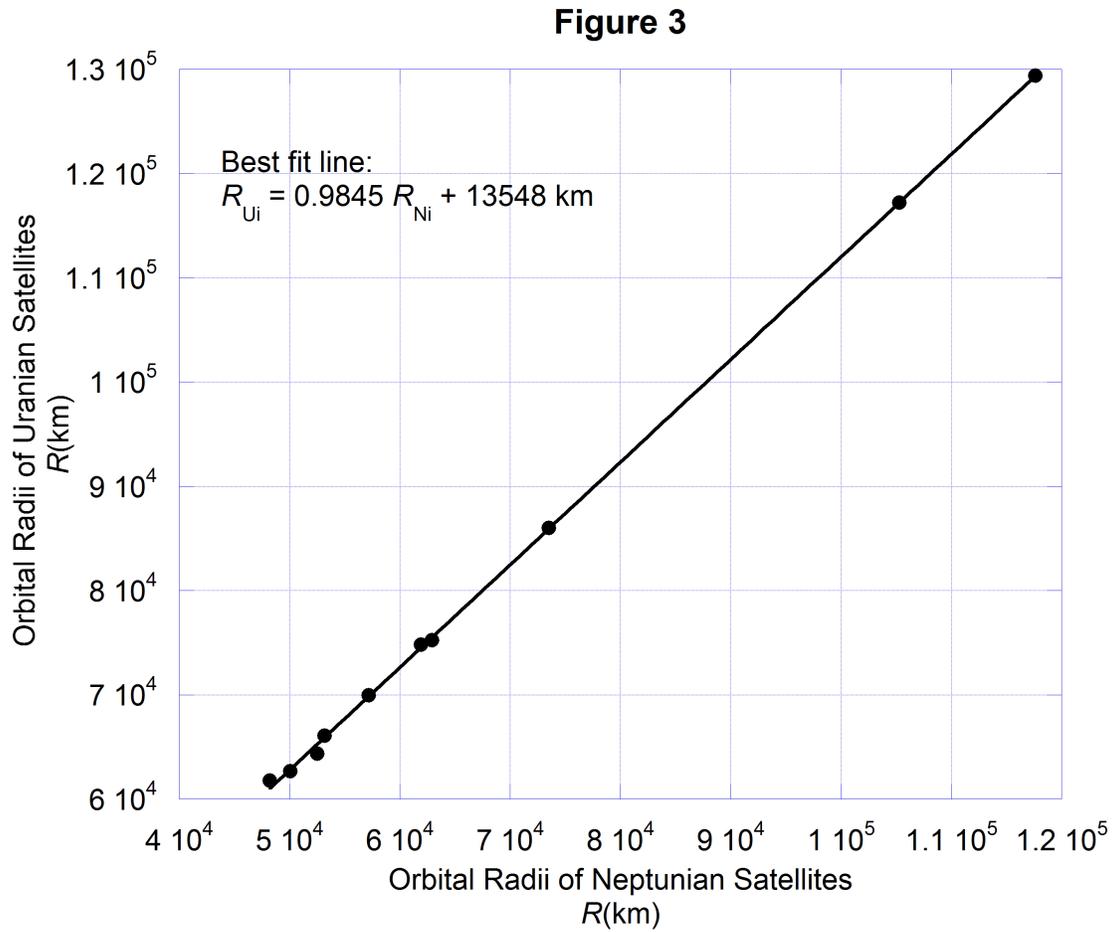

Figure 4.  The overlapping photon distributions for Uranus, and Saturn.  The photon energies and orbital radii for the uranian satellites are shifted by -3090 cm⁻¹ and 46700 km respectively to achieve overlap.

**Figure 4**

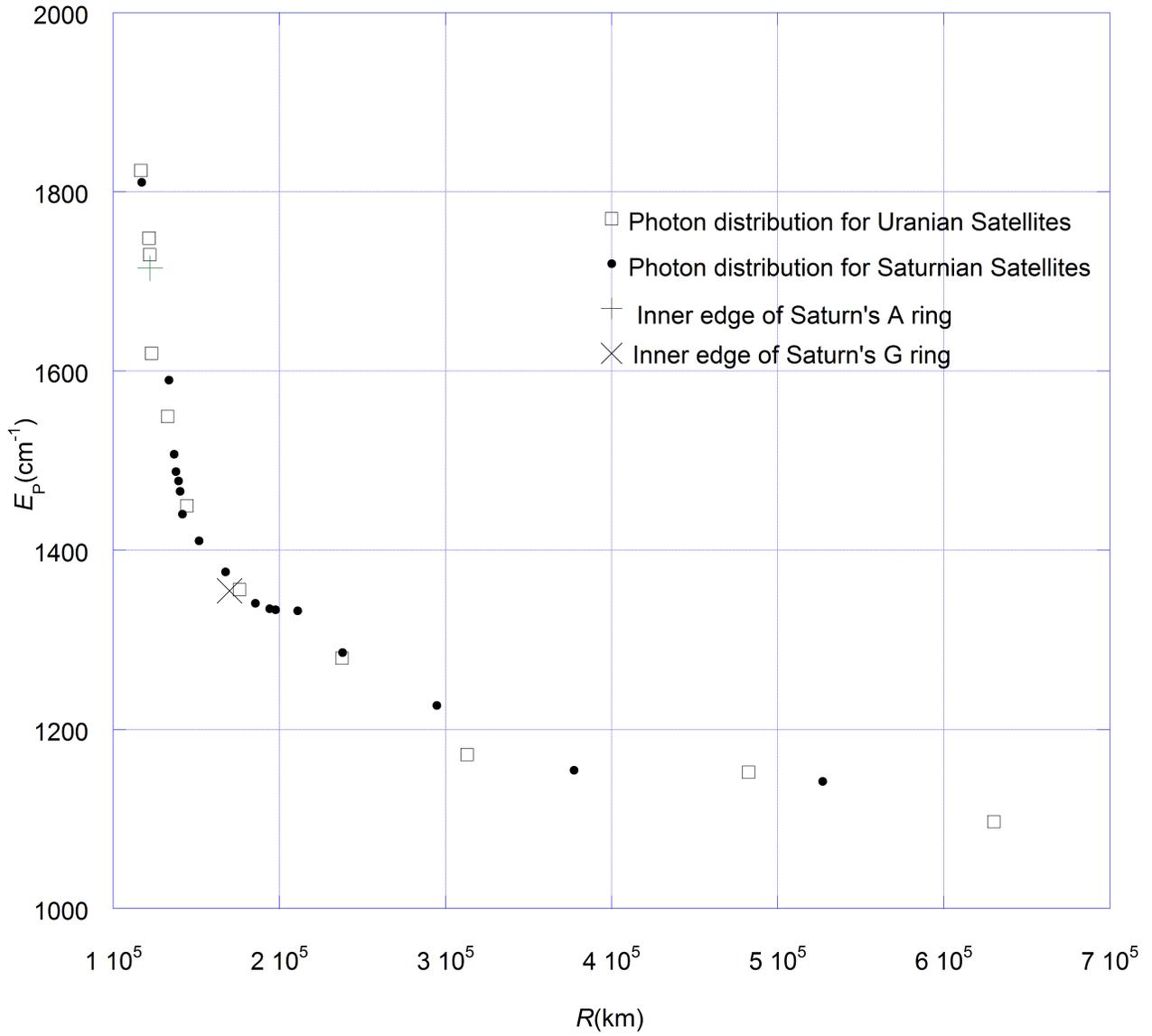

Figure 5.  Comparison of trial TD's with a calculated (Mousis 2003) temperature distribution in Uranus's protosatellite disk $10^4$ years after the giant impact occurred.  The trial TD's are calculated with $E_b = 2500$ cm$^{-1}$ (filled circles) and $E_b = 2700$ cm$^{-1}$ (open circles).   $E_b = 2500$ cm$^{-1}$ is used in subsequent calculations.

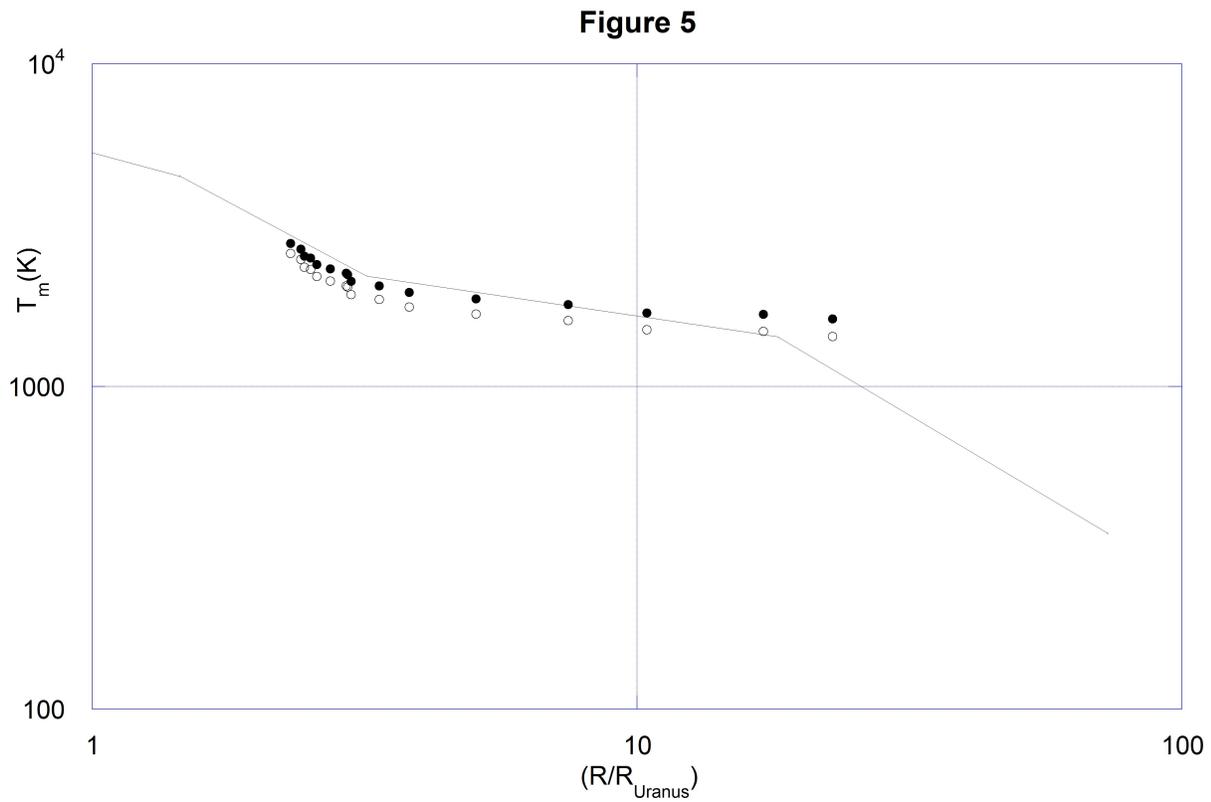

Figure 6. The composite TD of Uranus, Jupiter and Neptune on (a) a linear scale and (b) a log-log scale for comparison with Figure 6 in Alibert (2005).

## Figure 6a

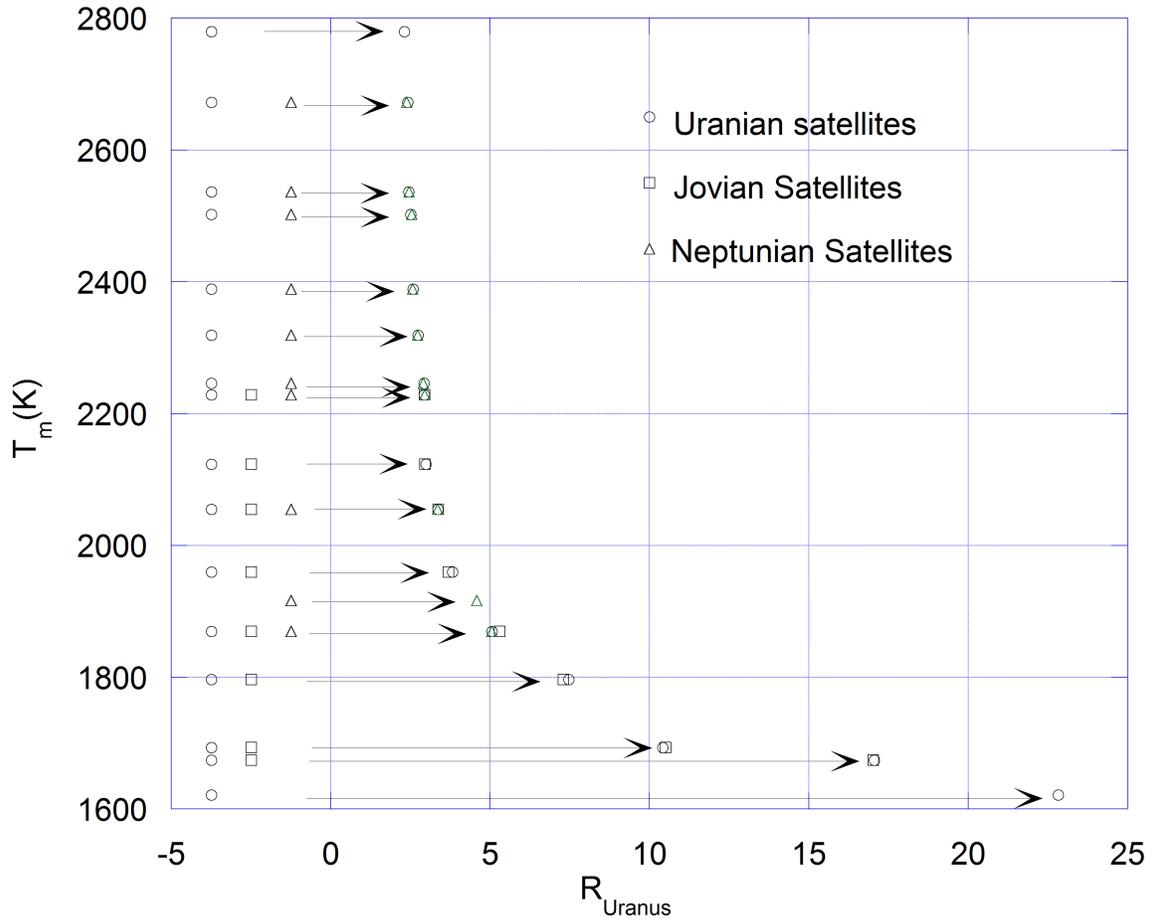

## Figure 6b

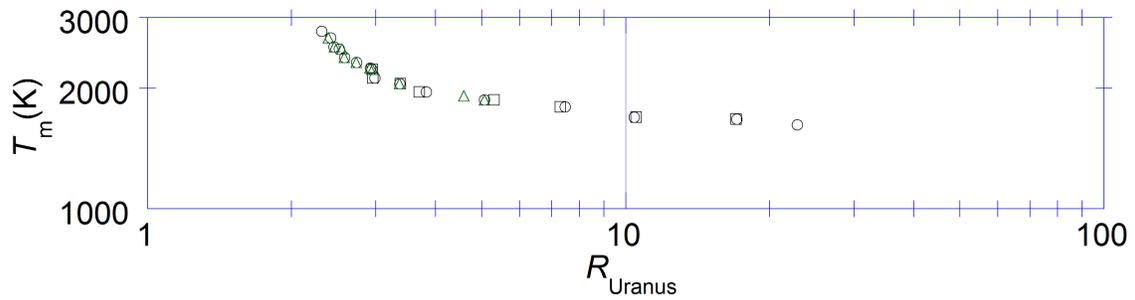

Figure 7. A logarithmic line graph of orbital radii of uranian satellites. The plotted points have widths that are large enough for most points to overlap. Two of the spaces on the line indicate where the slope of the TD is zero. One is the large space on the right between satellites Ophelia and Bianco. The other is on the left between Ring 4 and Ring Alpha. The three spaces on the right correspond to a broad peak and the space on the left to a narrower valley.

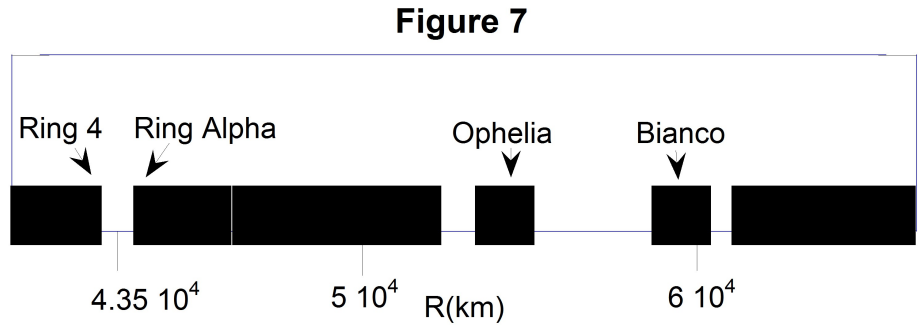

Figure 8. The complete TD for Uranus. Orbital radii and midplane temperatures are taken from Tables 4 and 5.

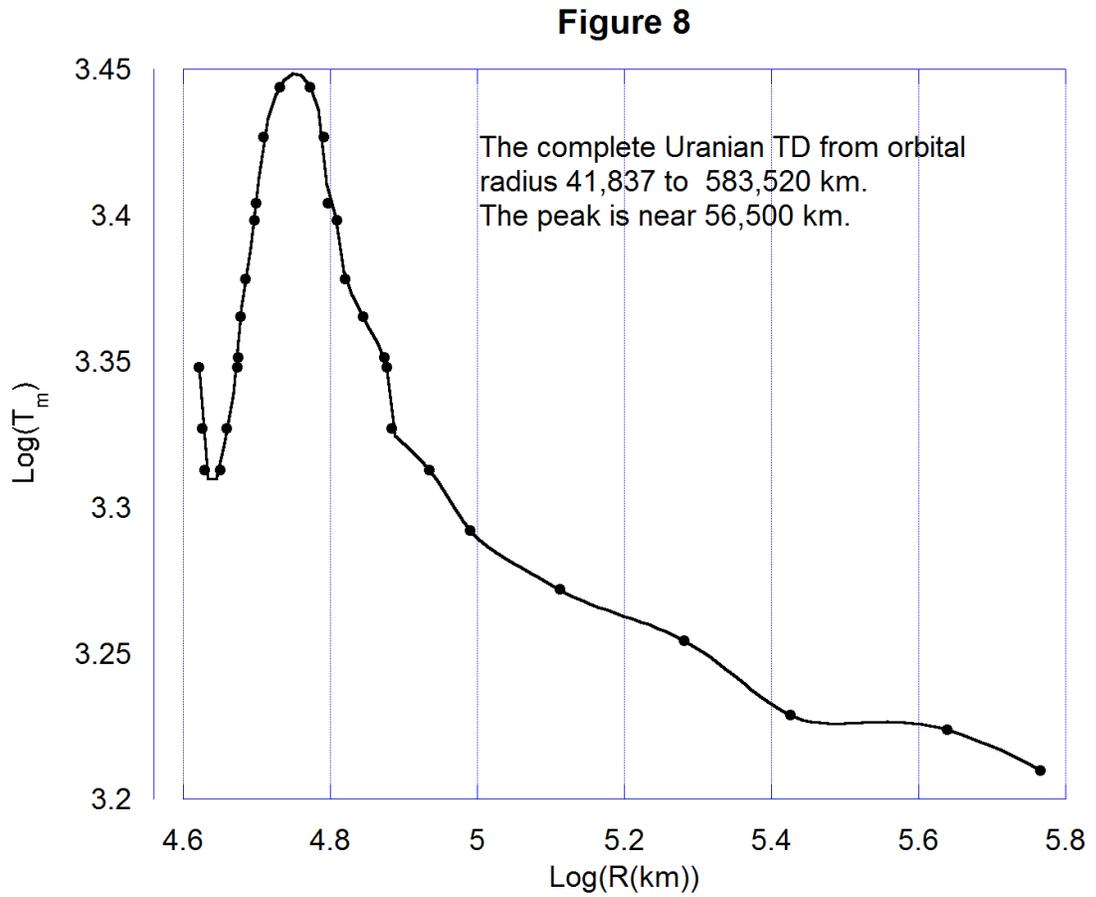

**Figure 8**

The complete Uranian TD from orbital radius 41,837 to 583,520 km. The peak is near 56,500 km.

Figure 9. Temperature distributions for the solar nebula at the onset and end of a downward transition (Lin and Papaloizou 1985).

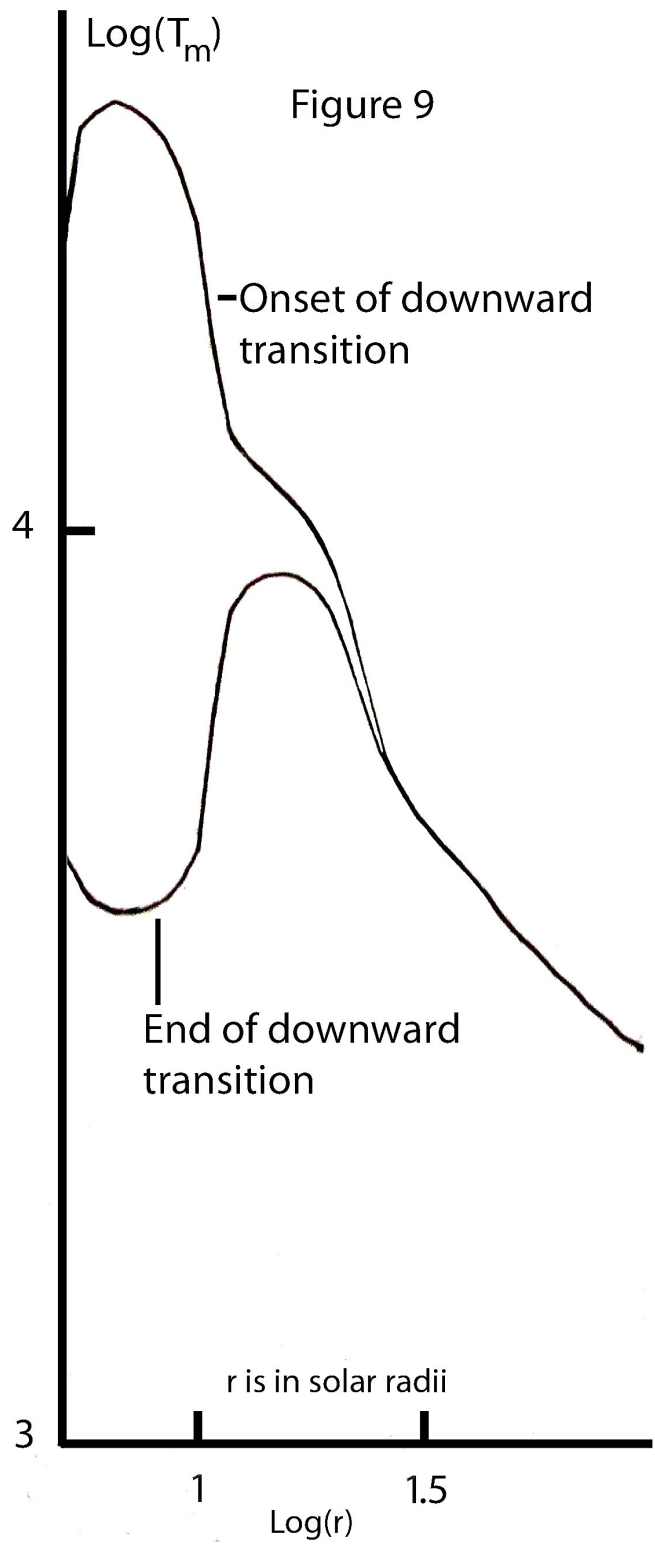

Figure 10.  The fit of Saturn's TD to a temperature distribution in Saturn's subnebula late in its
evolution (solid line) (Mousis et al. 2002).   The data for the TD (filled circles) is taken from
Table 3.  In the fitting process the $E_b$ value that yields the best fit is 945 cm$^{-1}$.

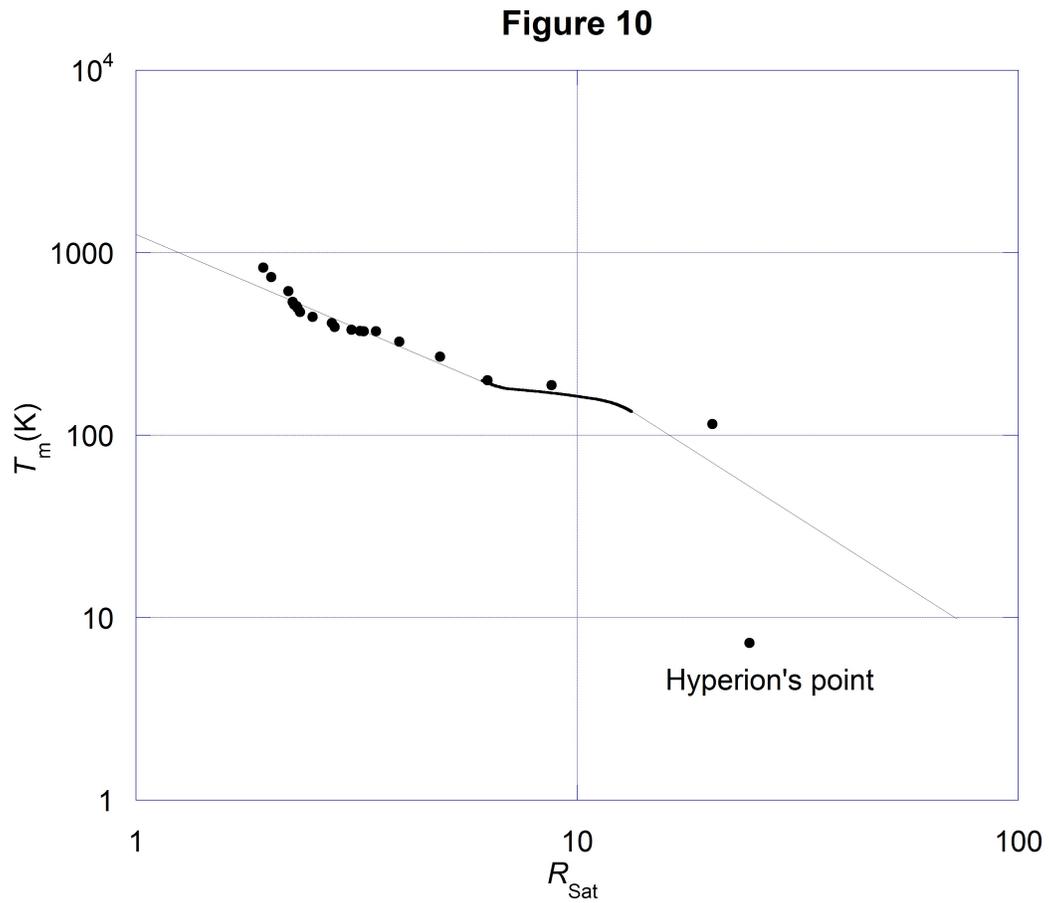

Figure 11. The composite TD of Uranus, Jupiter, Neptune and Saturn. Data is taken from Tables 3, 4 and 5. The temperatures for the TD's of Uranus, Jupiter and Neptune are shifted downward by 1480 K and the orbital radii for Saturn's satellites are adjusted using Eqn. (13) to achieve overlap of the TD's

**Figure 11**

TD's of Uranus, Jupiter and Neptune - filled circles
TD of Saturn - open circles

The numbers 2 or 3 next to some of the filled circles indicate where 2 or 3 points overlap.

$R_{\text{Uranus}}$

$T_\text{m}$(K)